\documentclass[showpacs,aps,prl,twocolumn,superscriptaddress]{revtex4}

\usepackage{graphicx}% Include figure files
\usepackage{dcolumn}% Align table columns on decimal point
\usepackage{bm}% bold math

\usepackage{color}

\newcommand{\ee}{$e^{+}e^{-}$}
\newcommand{\mm}{$\mu^{+}\mu^{-}$}
\newcommand{\kk}{$K^{+}K^{-}$}
\newcommand{\mev}{MeV/$c^2$}
\newcommand{\gev}{GeV/$c^2$}
\newcommand{\phiee}{$\phi \rightarrow e^{+}e^{-}$}

\newcommand{\phikk}{$\phi \rightarrow K^{+}K^{-}$}

\newcommand{\gammaphiee}{$\Gamma_{\phi\to e^+e^-}$}
\newcommand{\gammaphikk}{$\Gamma_{\phi\to K^+K^-}$}
\newcommand{\nphiee}{$N_{\phi\to e^+e^-}$}
\newcommand{\nphikk}{$N_{\phi\to K^+K^-}$}
\newcommand{\alphaphiee}{$\alpha_{\phi\to e^+e^-}$}
\newcommand{\alphaphikk}{$\alpha_{\phi\to K^+K^-}$}

\newcommand{\gammaphill}{$\Gamma_{\phi\to l^+l^-}$}
\newcommand{\gammaphikkb}{$\Gamma_{\phi\to K \bar K}$}

\begin{document}

%*******************************************************%
% title
%*******************************************************%
\title{Nuclear-matter modification of decay widths
in the $\phi \to e^+ e^- $ and $\phi \to K^+ K^-$ channels}

%*******************************************************%
% author lists
%*******************************************************%
\newcommand{\kyoto}{Department of Physics, Kyoto University,
 Kitashirakawa Sakyo-ku, Kyoto 606-8502, Japan}
\newcommand{\kek}{Institute of Particle and Nuclear Studies, KEK, 1-1
 Oho, Tsukuba, Ibaraki 305-0801, Japan}
\newcommand{\riken}{RIKEN, 2-1 Hirosawa, Wako, Saitama 351-0198, Japan}
\newcommand{\cns}{Center for Nuclear Study, Graduate School of Science,
 University of Tokyo, 7-3-1 Hongo, Tokyo 113-0033, Japan}
\newcommand{\rika}{Faculty of Science and Technology, Tokyo University
 of Science, 2641 Yamazaki, Noda, Chiba 278-8510, Japan}
\newcommand{\icepp}{ICEPP, University of Tokyo, 7-3-1 Hongo, Tokyo
 113-0033, Japan}
\newcommand{\tohoku}{Physics Department, Graduate School of Science,
 Tohoku University, Sendai 980-8578, Japan}
\newcommand{\tokyo}{Department of Physics, University of Tokyo, 7-3-1
 Hongo, Tokyo 113-0033, Japan}

\affiliation{\kyoto}
\affiliation{\kek}
\affiliation{\riken}
\affiliation{\cns}

\author{F.~Sakuma} \email{sakuma@nh.scphys.kyoto-u.ac.jp}
 \affiliation{\kyoto}
\author{J.~Chiba} \altaffiliation[Present Address: ]{\rika}
 \affiliation{\kek}
\author{H.~En'yo} \affiliation{\riken}
\author{Y.~Fukao} \affiliation{\kyoto}
\author{H.~Funahashi} \affiliation{\kyoto}
\author{H.~Hamagaki}\affiliation{\cns}
\author{M.~Ieiri} \affiliation{\kek}
\author{M.~Ishino} \altaffiliation[Present Address: ]{\icepp}
 \affiliation{\kyoto}
\author{H.~Kanda} \altaffiliation[Present Address: ]{\tohoku}
 \affiliation{\kyoto}
\author{M.~Kitaguchi} \affiliation{\kyoto}
\author{S.~Mihara} \altaffiliation[Present Address: ]{\icepp}
 \affiliation{\kyoto}
\author{K.~Miwa} \affiliation{\kyoto}
\author{T.~Miyashita} \affiliation{\kyoto}
\author{T.~Murakami} \affiliation{\kyoto}
\author{R.~Muto} \affiliation{\riken}
\author{T.~Nakura} \affiliation{\kyoto}
\author{M.~Naruki} \affiliation{\riken}
\author{K.~Ozawa} \altaffiliation[Present Address: ]{\tokyo}
 \affiliation{\cns}
\author{O.~Sasaki} \affiliation{\kek}
\author{M.~Sekimoto} \affiliation{\kek}
\author{T.~Tabaru} \affiliation{\riken}
\author{K.~H. Tanaka} \affiliation{\kek}
\author{M.~Togawa} \affiliation{\kyoto}
\author{S.~Yamada} \affiliation{\kyoto}
\author{S.~Yokkaichi} \affiliation{\riken}
\author{Y.~Yoshimura} \affiliation{\kyoto}

\collaboration{KEK-PS E325 Collaboration} \noaffiliation

%*******************************************************%
% date
%*******************************************************%
\date{\today}

%*******************************************************%
% abstract
%*******************************************************%
\begin{abstract}
The invariant mass spectra of \phikk\ are measured in 12 GeV $p+A$
 reactions in order to search for the in-medium modification of $\phi$
 mesons.
The observed \kk\ spectra are well reproduced by the relativistic
 Breit-Wigner function with a combinatorial background shape in three
 $\beta\gamma$ regions between 1.0 and 3.5.
The nuclear mass number dependence of the yields of the \kk\ decay
 channel is compared to the simultaneously measured \ee\ decay channel
 for carbon and copper targets.
We parameterize the production yields as 
 $\sigma \left( A \right) = \sigma_0 \times A^\alpha$
 and obtain \alphaphikk $-$ \alphaphiee\ to be 0.14~$\pm$~0.12.
Limits are obtained for the partial decay widths of the $\phi$ mesons in
 nuclear matter.
\end{abstract}

%*******************************************************%
% PACS numbers
%*******************************************************%
\pacs{25.40.Ve, 14.40.Cs, 21.65.+f, 25.75.Dw}
%25.40.Ve = Other reactions above meson production thresholds
%           (energies>400 MeV)
%14.40.Cs = Other mesons with S=C=0, mass<2.5 GeV
%21.65.+f = Nuclear matter
%25.75.Dw = Particle and resonance production

\maketitle

%*******************************************************%
% Introduction
%*******************************************************%
The properties of hadrons, such as mass, decay width, and branching
 ratio, have been extensively studied and well established in the
 history of particle physics.
Recent interests have been extended to understanding how these
 properties are modified in hot or dense matter.
This issue is of fundamental importance since such modifications can be
 related to the basic nature of QCD, spontaneous chiral symmetry
 breaking, i.e., the mechanism that creates most of the hadron masses.
Inspired by many theoretical works related to this subject, several
 experiments including ours have been carried out.

Amongst the many types of hadrons, the $\phi$ meson, which is a vector
 meson (1$^{- -}$) of an almost pure $s \bar s$ state, has very
 attractive features for use as a probe to detect the possible changes
 in its properties.
Its natural width is narrow ($\Gamma_\phi$ = 4.26 \mev)~\cite{PDG04}
 without any nearby resonance; therefore, we may
 be able to clearly detect the possible mass modification.
Theoretically, various models predict the in-medium mass modification of
 the $\phi$ meson both in dense and hot
 matter~\cite{HL92,KWW98,OR00,CV03,BiR90,BG91,SH97,PKL02}.
The predicted decrease in mass at normal nuclear density is up to 40 \mev\
 and the width broadening is up to 45 \mev.
Moreover, since $m_\phi$ is only 32 (24) \mev\ greater than $2m_{K^\pm}$
 ($2m_{K^0}$), the partial decay width \gammaphikkb\ is sensitive even
 to a small change in the spectral function of the $\phi$ meson and/or
 kaon.
Several theoretical models point out the possible change in the
 branching ratio \gammaphikkb/\gammaphill\ in a nuclear
 medium~\cite{BiR90,BG91,LS91}.

There are few experimental reports on the search for the in-medium
 modification of the $\phi$ meson.
With regard to the ratio of the partial decay width
 \gammaphikkb/\gammaphill, the experiments NA49 and NA50 at the CERN-SPS
 reported $\phi$-meson yields in the \kk\ and \mm\ channels,
 respectively.
There are discrepancies by factors ranging from 2 to 4 between these
 measurements~\cite{NA49phikk00,NA50phimm03,SPSphiem01}.
Recently, the CERES experiment at the CERN-SPS reported new results
 in the \phiee\ and \phikk\ measurements~\cite{CERESphiek05}.
The results in the \kk\ channels are in agreement with those of NA49,
 and both decay channels are consistent with each other.

The present experiment E325, performed using the KEK 12 GeV proton
 synchrotron, recently reported the spectral modification of the $\phi$
 meson in nuclear matter measured in the \ee\ decay channel for the
 first time~\cite{MUTO05}.
We also reported the modification of $\rho$ and/or $\omega$ mesons in
 Refs.~\cite{OZAWA01,NARUKI06}.
In the present study, new results are reported on the shape analysis
 for the \phikk\ invariant mass spectra, and the nuclear mass number
 dependence of $\phi$-meson production is compared between the \ee\ and
 \kk\ decay channels to determine whether the ratio of the partial decay
 width \gammaphikk/\gammaphiee\ depends on the nuclear size.
This work is an advanced study over Ref.~\cite{TABARU06}, where we
 reported the production cross-sections  and their rapidity and
 transverse momentum dependences of the $\omega$  and $\phi$ mesons
 measured in the present experiment. 

The cross section for a nuclear target of mass number $A$ is
 parameterized as 
$\sigma \left( A \right) = \sigma_0 \times A^\alpha$.
When the $\phi$ meson and/or kaon is modified in a medium and
 \gammaphikk/\gammaphiee\ changes, the ratio of the $\phi$-meson yield
 $R$ = \nphikk/\nphiee\ becomes dependent on the mass number since a
 larger number of $\phi$ mesons are to be modified in a larger nucleus.
Consequently, by using two different nuclear targets $A_1$ and $A_2$,
 the difference in the $\alpha$ parameter between \phiee\
 (\alphaphiee) and \phikk\ (\alphaphikk) can be related to $R$ as
 follows:
\begin{eqnarray}
 \Delta \alpha  & = & \alpha _{\phi  \to K^ +  K^ -  }  - \alpha _{\phi  \to
  e^ +  e^ -  }  \nonumber \\ 
  & = & \frac{{\ln \left[ {\frac{{N_{\phi  \to K^ +  K^ -  } \left( {A_1 }
							 \right)}}{{N_{\phi  \to K^ +  K^ -  } \left( {A_2 } \right)}}} \right] - \ln \left[ {\frac{{N_{\phi  \to e^ +  e^ -  } \left( {A_1 } \right)}}{{N_{\phi  \to e^ +  e^ -  } \left( {A_2 } \right)}}} \right]}}{{\ln \left( {{{A_1 } \mathord{\left/
 {\vphantom {{A_1 } {A_2 }}} \right.
 \kern-\nulldelimiterspace} {A_2 }}} \right)}} \nonumber \\ 
  & = & {{\ln \left[ {{{R\left( {A_1 } \right)} \mathord{\left/
 {\vphantom {{R\left( {A_1 } \right)} {R\left( {A_2 } \right)}}} \right.
 \kern-\nulldelimiterspace} {R\left( {A_2 } \right)}}} \right]}
  \mathord{\left/ {\vphantom {{\ln \left[ {{{R\left( {A_1 } \right)}
				    \mathord{\left/
 {\vphantom {{R\left( {A_1 } \right)} {R\left( {A_2 } \right)}}} \right.
 \kern-\nulldelimiterspace} {R\left( {A_2 } \right)}}} \right]} {\ln
  \left( {{{A_1 } \mathord{\left/
 {\vphantom {{A_1 } {A_2 }}} \right.
 \kern-\nulldelimiterspace} {A_2 }}} \right)}}} \right.
 \kern-\nulldelimiterspace} {\ln \left( {{{A_1 } \mathord{\left/
 {\vphantom {{A_1 } {A_2 }}} \right.
 \kern-\nulldelimiterspace} {A_2 }}} \right)}}.
\end{eqnarray}
It is important that most of the experimental effects are canceled in
 obtaining $\alpha$.

The details of the experiment can be found elsewhere~\cite{SEKIMOTO04}.
A primary proton beam of 12 GeV with a typical intensity of 10$^9$
 protons per 1.8-s spill was focused on carbon and copper targets, and
 the spectrometer detected \ee\ and \kk\ pairs simultaneously.
In the present article, we used the \ee-triggered data collected in
 2001 and 2002 and the \kk-triggered data collected in 2001.
In 2001, one carbon (0.092~g/cm$^2$) and two copper targets
(0.073~g/cm$^2$ each) were used,
 while in 2002, one carbon (0.184~g/cm$^2$) and four copper targets
 (0.073~g/cm$^2$ each) were used.
The targets were aligned in a line so that the same beam normalization
 could be used for all the targets.

%*******************************************************%
% Results
%*******************************************************%
\begin{figure}
 \includegraphics[width=8.6cm]{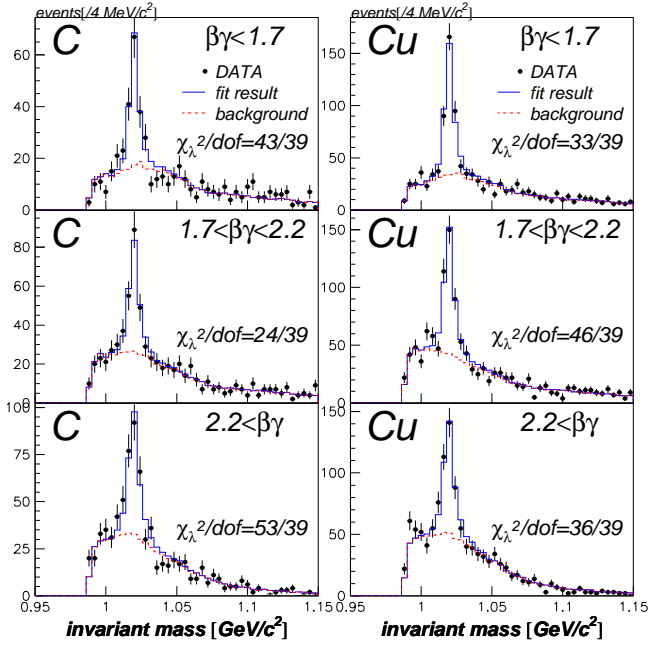}
 \caption{Invariant mass distributions of \kk\ pairs.
 The closed circles represent the observed distributions, and the solid
 lines represent the fit results with the expected \phikk\ shape on the
 combinatorial background shown with dashed lines.
 \label{fig:kkmass}}
\end{figure}

%-------------------------------------------------------%
% phi->KK Shapes
%-------------------------------------------------------%
The invariant mass distributions of the \kk\ pairs are shown in
 Fig.~\ref{fig:kkmass}.
To examine the modification in the mass shape as a function of
 $\beta\gamma$ ($=p/m$), the data were divided into three $\beta\gamma$
 regions.
Each mass spectrum was fitted with a combinatorial background shape
\footnote{According to the {\scshape jam} results, the contribution
 from the correlated pairs in the background is negligible.
}
 and a resonance shape of \phikk\ in the mass range between $2m_{K^\pm}$
 and 1.15 \gev.
The fits were performed by using the maximum likelihood
 method~\cite{LIKELI83}; the ``likelihood chi-square $\chi_\lambda
 ^2$'' is shown in Fig.~\ref{fig:kkmass}.
The combinatorial background shape was evaluated by the event-mixing
 method.
For the $\phi$-meson resonance shape, the relativistic Breit-Wigner
 (RBW) distribution with the natural mass ($m_0$) and
 decay width ($\Gamma_0$) was used after taking the spectrometer
 performance into account.
The RBW distribution is given by
\begin{equation}
\frac{{d\sigma }}{{dm}} =
 \frac{{m m_0  \Gamma \left( m \right)}}{{\left(
     {m^2  - m_0  ^2 } \right)^2  + \left( {m_0  \Gamma \left( m
            \right)} \right)^2 }} ,
\end{equation}
 where
$\Gamma \left( m \right) = \Gamma _0 \left( {{q \mathord{\left/
 {\vphantom {q {q_0 }}} \right.
 \kern-\nulldelimiterspace} {q_0 }}} \right)^3 \left( {{{m_0 }
 \mathord{\left/
 {\vphantom {{m_0 } m}} \right.
 \kern-\nulldelimiterspace} m}} \right)$,
$q = \left( {{{m^2 } \mathord{\left/
 {\vphantom {{m^2 } 4}} \right.
 \kern-\nulldelimiterspace} 4} - m_{K^\pm}^2 } \right)^{{1 \mathord{\left/
 {\vphantom {1 2}} \right.
 \kern-\nulldelimiterspace} 2}}, $
and
$q_0  = \left( {{{m_0^2 } \mathord{\left/
 {\vphantom {{m_0^2 } 4}} \right.
 \kern-\nulldelimiterspace} 4} - m_{K^\pm}^2 } \right)^{{1 \mathord{\left/
 {\vphantom {1 2}} \right.
 \kern-\nulldelimiterspace} 2}} $~\cite{JACKSON64}.

\begin{figure}
 \includegraphics[width=8.6cm]{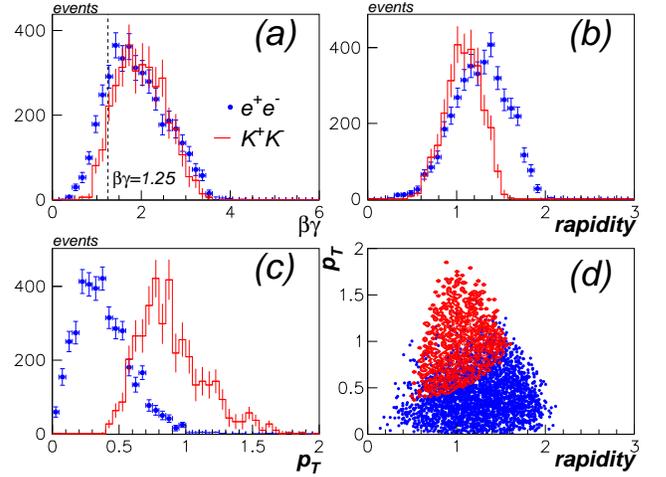}
 \caption{
 Observed kinematical distributions of \phiee\ and \phikk.
 Distributions of (a) $\beta\gamma$, (b) rapidity, (c) $p_T$, and (d)
 $p_T$ and rapidity.
 The points represent \phiee, and the lines represent \phikk.
 The histograms for \phikk\ are scaled by a factor of $\sim$3.
 \label{fig:eekkparm}}
\end{figure}

The spectrometer performance was examined by a detailed
 detector simulation using the {\scshape geant4} toolkit~\cite{GEANT02}.
The kinematical distributions of the $\phi$ mesons used in the
 simulation were obtained from the nuclear cascade code {\scshape
 jam}~\cite{JAM99}, which reproduced our measured distributions fairly
 well~\cite{TABARU06}.
From the data, we obtained the peak position $m$ = 1019.43 $\pm$ 0.21 (stat)
 $\pm$ 0.04 (syst) \mev\ and the mass resolution $\sigma$ = 1.91 $\pm$
 0.29 (stat) $\pm$ 0.23 (syst) \mev\ for the copper targets, while the
 simulation gave $m$ = 1019.49 \mev\ and $\sigma$ = 2.24 \mev.
The agreements are significantly remarkable, and we have used the
 simulated mass shape for the fits in Fig.~\ref{fig:kkmass}.

The observed \kk\ spectra are well reproduced by the fit in all
 the $\beta\gamma$ bins.
Therefore, the changes in mass spectra are not statistically significant
 in the \phikk\ channel.
Indeed, we have observed a mass shape modification in the \phiee\
 channel~\cite{MUTO05} in the very low $\beta\gamma$ region
 ($\beta\gamma<1.25$) for the copper target data.
In Fig.~\ref{fig:eekkparm}, we compare the acceptances of the \kk\ and
 \ee\ channels.
In the region $\beta\gamma<1.25$, we have very limited statistics for
 \phikk, and we cannot obtain a reasonable fit for the \kk\ data.
Thus, it is impossible to compare the \phikk\ shape directly with the
 modified shape observed in the \ee\ channel.
It can be concluded that in the region $\beta\gamma>1.25$,
 both \phikk\ and \phiee\ do not show signs of shape
 modification.

%-------------------------------------------------------%
% Nuclear Dependences
%-------------------------------------------------------%
\begin{table}
 \caption{$\phi$-meson yields and the $\alpha$ parameters for the
 \ee\ and \kk\ decay channels evaluated by the fits.
 The first errors are statistical and the second are systematic.
 \label{tab:nphi}}
 \begin{ruledtabular}
  \begin{tabular}{c@{}c@{}c@{}c}
   $\beta\gamma$ & $N_\phi$(C) & $N_\phi$(Cu) & $\alpha$ \\\hline
   \phiee & & & \\
   0--1.25    & 331$\pm$32$\pm$6  & 703$\pm$50$\pm$26  & 1.18$\pm$0.07$\pm$0.02 \\
   1.25--1.75 & 702$\pm$51$\pm$9  & 899$\pm$63$\pm$4   & 0.87$\pm$0.06$\pm$0.02 \\
   1.75--2.5  & 945$\pm$62$\pm$14 & 1096$\pm$71$\pm$12 & 0.81$\pm$0.06$\pm$0.01 \\
   2.5--3.5   & 579$\pm$52$\pm$4  & 626$\pm$56$\pm$5   & 0.77$\pm$0.08$\pm$0.01 \\
   \hline
   \phikk & & & \\
   1--1.7   & 99$\pm$20$\pm$18  & 285$\pm$29$\pm$27 & 1.38$\pm$0.13$\pm$0.07 \\
   1.7--2.2 & 143$\pm$24$\pm$19 & 279$\pm$32$\pm$30 & 1.14$\pm$0.12$\pm$0.03 \\
   2.2--3.5 & 177$\pm$26$\pm$11 & 269$\pm$33$\pm$21 & 0.99$\pm$0.12$\pm$0.03 \\
  \end{tabular}
 \end{ruledtabular}
\end{table}

We obtained the $\phi$-meson yields for three bins for \kk\ and four bins for
 \ee\ as functions of $\beta\gamma$.
For the \kk\ decay channel, the $\phi$-meson yields were obtained by
 integrating from  $2m_{K^\pm}$ to 1.07 \gev\ after subtracting the
 background described before.
To obtain the $\phi$-meson yields for the \ee\ decay channel, we
 employed the same procedure as used in Ref.~\cite{MUTO05}; in that
 study, the spectra were fitted with a simulated $\phi$ resonance shape
 and a quadratic background curve.
The $\phi$-meson yields were then obtained by integrating the data
 in the mass range between 0.9 and 1.1 \gev\, after subtracting
 the background.
Thus, the excess yield observed in Ref.~\cite{MUTO05} must also be
 counted as $\phi$ mesons.

Table~\ref{tab:nphi} summarizes the yields for both decay channels with
 the systematic errors containing the contributions from the
 uncertainty in the background estimation
\footnote{Larger values of \nphiee\ as compared to Ref.~\cite{MUTO05} were
 obtained with looser analysis cuts, which were optimized to determine 
 \alphaphiee($y,p_T$).}.
In Table~\ref{tab:nphi}, the $\alpha$ parameters are corrected for the
 target dependence of the experimental efficiencies.
The most significant correction was the difference in the geometrical
 acceptance for each target position.
The effect was estimated by the {\scshape jam} simulation, and $\alpha$
 changed by a maximum of 0.04.
The other corrections due to the trigger, tracking, and analysis were
 estimated to be small and were included in the systematic errors
 together with those from the background estimations.
In the analysis thereafter, the errors in $\alpha$ are the quadratic
 sums of the statistical and systematic errors.

\begin{figure}
 \includegraphics[width=8.6cm]{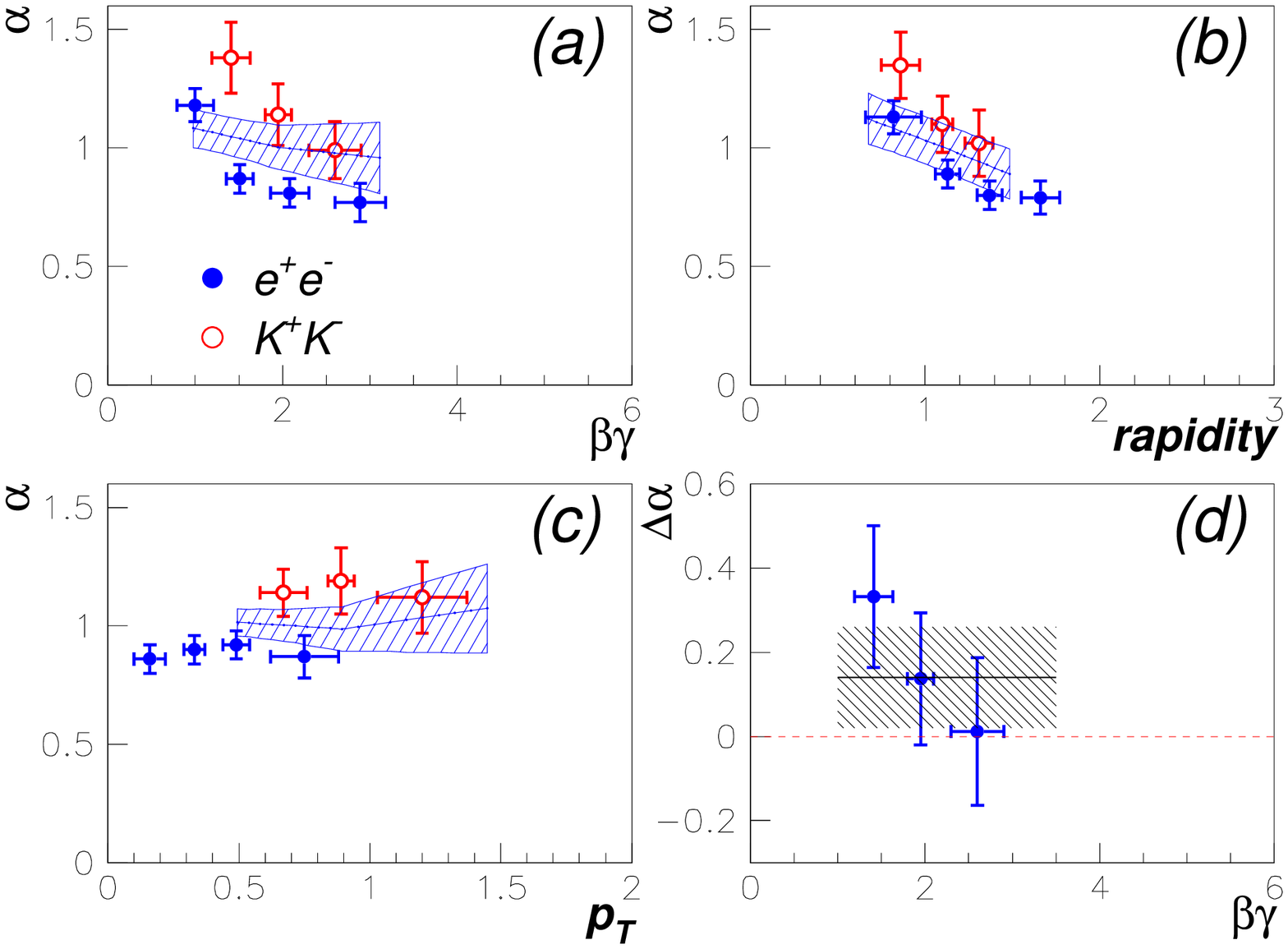}
 \caption{
 $\alpha$ parameters of \phiee\ (closed circles) and \phikk\
 (open circles) as functions of (a) $\beta\gamma$, (b) rapidity,  and (c)
 $p_T$.
 The horizontal error bars are the RMS values of the horizontal bins.
 The hatched bands show corrected \alphaphiee\ to be compared with
 \alphaphikk\ (see text).
 In (d), the differences between \alphaphikk\ and \alphaphiee\ in the
 kaon acceptance window are shown.
 The average value and error are also plotted as a hatched band.
 \label{fig:alpha}}
\end{figure}

The observed $\alpha$ parameters are plotted as functions of
 $\beta\gamma$ in Fig.~\ref{fig:alpha}(a) and as functions of
 the rapidity ($y$) and transverse momentum ($p_T$) in (b) and (c).
The \ee\ and \kk\ decay channels cannot be compared directly because of
 the difference in the detector acceptance between the \ee\
 and \kk\ decay channels, as shown in Fig.~\ref{fig:eekkparm}.
Therefore, we determined \alphaphiee\ two-dimensionally on the
 $y$-$p_T$ plane under the assumption that \alphaphiee\ is linearly
 dependent on $y$ and $p_T$, and estimated the values corresponding to
 the kaon acceptance windows
\footnote{Indeed, in our detector acceptances, {\scshape jam} predicts
 such a dependence of the $\alpha$ parameter in the $y$-$p_T$ plane.}.
To determine the plane \alphaphiee($y,p_T$), the \ee\ data were divided
 into $3 \times 3$ bins in the $y$-$p_T$ plane, and they were fitted
 with the function $\alpha (y,p_T) = a \times y + b \times p_T + c$.
We obtained a = $-$0.32 $\pm$ 0.11, b = 0.13 $\pm$ 0.17, c = 1.24 $\pm$
 0.15, and $\chi^2$/dof = 4.2/6, indicating that the above assumption is
 statistically acceptable.

In Fig.~\ref{fig:alpha}(a), we show the estimated values of
 \alphaphiee($\beta\gamma$) in the kaon acceptance window as a hatched
 band; this can be compared to the measured value of \alphaphikk.
The difference $\Delta \alpha$ in the kaon
 acceptance is plotted in Fig.~\ref{fig:alpha}(d).
We expect that $\Delta \alpha$ should be zero when
 \gammaphikk/\gammaphiee\ does not change in nuclear media.
Although it is interesting to observe that $\Delta \alpha$ increases
 when $\beta\gamma$ decreases, $\Delta \alpha$ at the lowest
 $\beta\gamma$ bin is only 0.33 $\pm$ 0.17; the average value is 0.14
 $\pm$ 0.12.
Therefore, \alphaphiee\ and \alphaphikk\ are statistically the same in
 the measured kinematic region.

%*******************************************************%
% Discussions
%*******************************************************%
On the basis of these results, the possible modification of the decay
 widths is discussed below.
In a medium, we expect the total and partial decay widths $\Gamma$ to change
 to $\Gamma^*$ according to the relation
\begin{eqnarray}
 {{\Gamma _\phi ^* } \mathord{\left/
 {\vphantom {{\Gamma _\phi ^* } {\Gamma _\phi ^0 }}} \right.
 \kern-\nulldelimiterspace} {\Gamma _\phi ^0 }} & = & 1 + k_{\text{tot}} \left(
							       {{\rho
							       \mathord{\left/
 {\vphantom {\rho  {\rho _0 }}} \right.
 \kern-\nulldelimiterspace} {\rho _0 }}} \right), \nonumber \\ 
 {{\Gamma _{\phi  \to K^ +  K^ -  }^* } \mathord{\left/
 {\vphantom {{\Gamma _{\phi  \to K^ +  K^ -  }^* } {\Gamma _{\phi  \to
 K^ +  K^ -  }^0 }}} \right.
 \kern-\nulldelimiterspace} {\Gamma _{\phi  \to K^ +  K^ -  }^0 }} & = & 1 +
 k_K \left( {{\rho  \mathord{\left/
 {\vphantom {\rho  {\rho _0 }}} \right.
 \kern-\nulldelimiterspace} {\rho _0 }}} \right), \\ 
 {{\Gamma _{\phi  \to e^ +  e^ -  }^* } \mathord{\left/
 {\vphantom {{\Gamma _{\phi  \to e^ +  e^ -  }^* } {\Gamma _{\phi  \to
 e^ +  e^ -  }^0 }}} \right.
 \kern-\nulldelimiterspace} {\Gamma _{\phi  \to e^ +  e^ -  }^0 }} & = & 1 +
 k_e \left( {{\rho  \mathord{\left/
 {\vphantom {\rho  {\rho _0 }}} \right.
 \kern-\nulldelimiterspace} {\rho _0 }}} \right), \nonumber
\end{eqnarray}
 where $\Gamma^0$ is the value in vacuum and $\rho_0$ is the normal
 nuclear density \footnote{This linear assumption is consistent with the
 theoretical calculation in Fig. 4 of Ref.~\cite{OR00} at the $\phi$
 mass (1020 \mev).}.
We expect $k_{\text{tot}} \simeq k_K$ since the $\phi$ meson mainly
 decays into $K \bar K$ as long as such decays are kinematically
 allowed.

\begin{figure}
 \includegraphics[width=8.6cm]{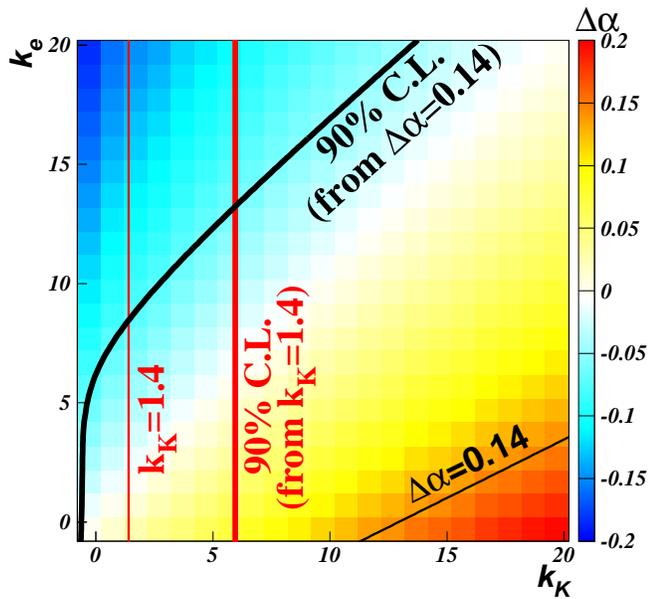}
 \caption{
 The obtained constraints on the in-medium modification of the
 partial decay widths of \phikk\ and \phiee\ (see text).
 \label{fig:da}}
\end{figure}

To estimate the relation between $\Delta \alpha$, $k_K$, and $k_e$,
 we performed the Monte-Carlo calculation described in
 Ref.~\cite{MUTO05}, where $\phi$ mesons were uniformly produced in a
 nucleus and decayed according to the values of $k_K$ and $k_e$.
In the plane of $k_K$ and $k_e$, the expected values of $\Delta \alpha$
 were obtained.
The results are shown in Fig.~\ref{fig:da} as a colored contour, where
 the measured $\Delta \alpha$ and its error provide the 90\% confidence
 limit.

Next, we have investigated how the \kk\ spectra could provide
 constraints for $k_K$.
Since we observed a significant excess on the low-mass side of the
 $\phi$-meson peak in the \phiee\ channel~\cite{MUTO05}, we consider
 that a similar excess can exist in the \phikk\ spectra.
We thus reanalyzed the \phikk\ spectra with the same fitting procedure
 as described earlier, except that the mass range from $2m_{K^\pm}$ to
 1.01 \gev\ (0.01 \gev\ below the $\phi$ peak) was excluded from the
 fit.
Due to the statistical limitation, we combined all the
 kinematical bins into one.

This procedure gives the amount of excess $N_{\text{ex}}$ as a surplus
 over the $\phi$-meson peak and the background.
Since the \kk\ acceptance is a function of the invariant mass, we
 corrected $N_{\text{ex}}/N_{\phi}$ and obtained 
0.044 $\pm$ 0.037 (stat) $\pm$ 0.058 (syst) and 0.076 $\pm$
 0.025 (stat) $\pm$ 0.043 (syst) for carbon and copper targets,
 respectively.
%Although these surpluses approximate zero, they statistically limit the
Although these surpluses are close to zero, they statistically limit the
 number of modified $\phi$ mesons in the \kk\ spectra.
The ratio $N_{\text{ex}}/N_{\phi}$ can be considered as the ratio of the
 number of $\phi$ mesons decayed inside the nucleus to that outside it,
 i.e.,
 $N_{\text{ex}}/N_{\phi} = N_{\phi}^{\text{in}}/N_{\phi}^{\text{out}}$.

The relation between $N_{\phi}^{\text{in}}/N_{\phi}^{\text{out}}$ and
 $k_K$ was also obtained by the Monte-Carlo calculation.
We assumed that the decays inside the half-density radius of the
 Woods-Saxon distribution contribute to $N_{\phi}^{\text{in}}$.
By using the calculated relation, $k_K$ was obtained as 1.4 $\pm$ 1.1
 (stat) $\pm$ 2.1 (syst); this was the average value for carbon and
 copper targets.
This result is plotted in Fig.~\ref{fig:da}, together with the
 upper-limit value, 6.0, at the 90\% confidence level.
%together with the 90\% confidence limit.

The two constraints shown in Fig.~\ref{fig:da} are the first
 experimental limits assigned to the in-media broadening of the partial
 decay widths of the $\phi$ meson.
To obtain the limits, we have renormalized the probability distribution
 functions by eliminating an unphysical region corresponding to
 $\Gamma^*/\Gamma^0<0$.

%*******************************************************%
% Summary
%*******************************************************%
To conclude, the experiment KEK-PS E325 measured the $\phi$-meson
 production via the \ee\ and \kk\ decay channels in 12 GeV $p+A$
 reactions.
The observed \kk\ spectra are well reproduced by the relativistic
 Breit-Wigner distribution with a combinatorial background shape.
The nuclear mass number dependences of \phiee\ and \phikk\ are consistent.
We have obtained the limits on the in-media decay widths for both the
 decay channels.

%*******************************************************%
% Acknowledgments
%*******************************************************%
\begin{acknowledgments}
We gratefully acknowledge all the staff members of KEK-PS.
This study was partly funded by the Japan Society for the Promotion
 of Science and a Grant-in-Aid for Scientific Research.
We would like to thank the staff members of the
 RIKEN Super Combined Cluster System (RSCC) and RIKEN-CCJ.
\end{acknowledgments}

%*******************************************************%
% bibliography
%*******************************************************%
%\bibliography{sakuma.bib}

\end{document}